\documentclass[%
reprint,
amsmath,amssymb,
aps,
prb,
]{revtex4-2}
\usepackage{graphicx}
\usepackage{CJKutf8}
\usepackage{epstopdf}
\usepackage{subfigure}
\usepackage{pifont}
\usepackage{bm}
\usepackage{xcolor}
\usepackage[colorlinks=true,linkcolor=blue,anchorcolor=blue,citecolor=blue,urlcolor=blue]{hyperref}
\begin{document}
\preprint{APS/123-QED}

\title{Effective perpendicular electric field as a probe for interlayer pairing in ambient-pressure superconducting La$_{2.85}$Pr$_{0.15}$Ni$_{2}$O$_{7}$ thin films}

\author{Junkang Huang$^{1}$}
\author{Tao Zhou$^{1,2}$}
\email{tzhou@scnu.edu.cn}

\affiliation{$^1$Guangdong Basic Research Center of Excellence for Structure and Fundamental Interactions of Matter, Guangdong Provincial Key Laboratory of Quantum Engineering and Quantum Materials, School of Physics, South China Normal University, Guangzhou 510006, China\\
	$^2$Guangdong-Hong Kong Joint Laboratory of Quantum Matter, Frontier Research Institute for Physics, South China Normal University, Guangzhou 510006, China}

\date{\today}

\begin{abstract}
Recent angle-resolved photoemission spectroscopy (ARPES) experiments on La$_{2.85}$Pr$_{0.15}$Ni$_2$O$_7$ thin films have revealed a superconducting gap near the diagonal direction for the $d_{x^2-y^2}$ orbital, confined to a narrow momentum region. Our numerical calculations indicate that interlayer pairing is dominant, resulting in an $s_\pm$-wave symmetry. We propose employing perpendicular electric fields, a practical method in thin film systems, to verify the pairing symmetry. Our calculations predict that such fields will induce Fermi arcs or nodal points and significantly modulate the gap near the diagonal, with the modulated region overlapping the area probed by ARPES. This approach provides an effective tool for probing pairing symmetry in bilayer nickelate thin films, especially given current limitations in experimental techniques.
\end{abstract}
\maketitle

\section{\label{Intro}Introduction}

A groundbreaking advancement in superconductivity research emerged with the discovery of high-temperature superconductivity in pressurized bilayer Ruddlesden-Popper nickelate La$_3$Ni$_2$O$_{7}$, exhibiting a superconducting transition temperature ($T_c$) of up to $80$~K~\cite{s41586-023-06408-7}. This milestone has sparked intense experimental and theoretical interest in nickel-based oxide superconductors, establishing them as a compelling platform for investigating the mechanisms of high-temperature superconductivity~\cite{Wang_2024,Liu2024,Zhang2024,PhysRevLett.132.256503,Dong2024,PhysRevX.14.011040,PhysRevLett.131.126001,PhysRevLett.131.236002,PhysRevB.108.165141,PhysRevB.108.L140505,PhysRevLett.134.076001,PhysRevB.111.104505,PhysRevB.110.094509,PhysRevB.109.165154,PhysRevB.109.L081105,arXiv2311.12769,PhysRevB.108.174511,xie20243221,s41467-024-53863-5,PhysRevB.108.L201108,arXiv2311.05491,PhysRevLett.132.146002,arXiv2402.07449,yang2024decom,PhysRevB.111.094504,arXiv2411.13554,PhysRevMaterials.8.124801,PhysRevB.108.174501,PhysRevB.110.024514,Huo2025,Wu2024,Luo2024}. Over the past year, analogous bilayer nickelate systems have demonstrated similarly remarkable properties: La$_2$PrNi$_2$O$_7$ was found to superconduct at $82.5$~K~\cite{Wang2024}, while La$_2$SmNi$_2$O$_{7-\delta}$ achieved an even higher $T_c$ of $91$~K under pressure~\cite{li2025ambient}. Notably, all these materials require external pressure to stabilize their superconducting states.

Recent breakthroughs have overcome this constraint by realizing ambient-pressure superconductivity in epitaxially stabilized thin-film nickelates. Specifically, La$_3$Ni$_2$O$_{7}$ and La$_{2.85}$Pr$_{0.15}$Ni$_2$O$_7$ thin films have exhibited superconducting transitions above 40~K under ambient conditions~\cite{Ko2025,Zhou2025}. These advances represent pivotal progress toward practical applications, bridging the gap between high-pressure bulk systems and ambient-pressure superconducting technologies.

Theoretical and experimental studies on La$_3$Ni$_2$O$_{7}$ and La$_{2.85}$Pr$_{0.15}$Ni$_2$O$_7$ thin-film superconductors reveal strong structural and electronic similarities with bilayer nickel-based bulk superconductors \cite{Ko2025,Zhou2025,arXiv2501.08022,arXiv2502.17831,arXiv2501.06875,arXiv2501.09255,arXiv2501.08204,wang2025electronic,arXiv2503.17223}. Both systems have tetragonal NiO$_2$ planes and
a nearly linear apical Ni-O-Ni bond angle, which may be a critical feature for their superconducting properties~\cite{Ko2025,Zhou2025}. Density functional theory (DFT) calculations further demonstrate that electronic states near the Fermi level in the thin films are primarily governed by Ni-$d_{x^2-y^2}$ and Ni-$d_{z^2}$ orbitals~\cite{arXiv2501.06875,arXiv2503.17223}, consistent with bulk ones~\cite{PhysRevLett.131.126001}.
Angle-resolved photoemission spectroscopy (ARPES) experiments have measured the superconducting gap magnitude near the diagonal direction of the $d_{x^2-y^2}$ orbital, revealing a nodeless gap with minimal magnitude in this direction~\cite{arXiv2502.17831}. This marks the first direct observation of the superconducting gap in bilayer nickelate superconductors, providing critical insight into their pairing mechanisms. However, current experimental limitations restrict measurements to an extremely narrow region of the Brillouin zone, potentially hindering comprehensive comparisons between theory and experiment~\cite{arXiv2502.17831}.

The Ruddlesden-Popper nickelate superconductors distinguish themselves from other unconventional high-temperature superconductors by the significant contribution of $d_{z^2}$ orbital electrons to the Fermi surface~\cite{PhysRevLett.131.126001,arXiv2501.06875,arXiv2503.17223}. This orbital, proposed to play a key role in superconductivity~\cite{s41586-023-06408-7,Huo2025,Wu2024}, mediates interlayer antiferromagnetic interactions that may promote interlayer electron pairing~\cite{xie20243221,s41467-024-53863-5,PhysRevB.108.L201108,PhysRevB.108.174511,arXiv2311.05491,PhysRevLett.132.146002,arXiv2402.07449,yang2024decom,PhysRevB.111.094504}. Recent scanning tunneling microscope (STM) experiments have revealed a two-gap structure, with a dominant interlayer pairing amplitude and a relatively smaller intralayer component~\cite{arXiv2506.01788}.

Previous studies on trilayer nickelate La$_4$Ni$_3$O$_{10}$ have shown that interlayer pairing can produce nodes or nodal lines in the superconducting order parameter, arising from the electronic inequivalence between the inner and outer NiO$_2$ layers~\cite{PhysRevB.110.L060506,Qiong2024}. In contrast, bilayer nickelate superconductors feature two electronically equivalent NiO$_2$ layers, so interlayer Cooper pairing in these systems yields a nodeless superconducting order parameter due to the symmetry of the pairing wavefunction. However, if this equivalence is experimentally disrupted, interlayer pairing could induce nodes or nodal lines—providing a critical test for the dominance of interlayer pairing in these materials.

It has been previously proposed that applying a perpendicular electric field to a La$_3$Ni$_2$O$_7$ thin film induces electron transfer from the top to the bottom layer, thereby breaking the equivalence between layers and strongly affecting interlayer pairing~\cite{arXiv2411.13554}. Motivated by these considerations, in this work we employ a two-orbital model tailored to La$_{2.85}$Pr$_{0.15}$Ni$_2$O$_7$ thin films and perform self-consistent mean-field calculations to investigate the superconducting pairing mechanism. By incorporating both interlayer and intralayer electron interactions, our results reveal that interlayer pairing dominates, yielding an effective $s_{\pm}$ symmetry. Upon introducing an effective electric field, we find that the pairing function near the diagonal direction for the $d_{x^2-y^2}$ orbital is significantly modified, with additional nodes being induced. This approach thus offers a powerful means for probing the pairing symmetry in nickelate thin films.

The remainder of this paper is organized as follows. In Section~II, we describe the model and elaborate on the formalism. In Sections~III, we present our self-consistent calculations without the perpendicular electric field. The numerical results in the presence of the perpendicular electric field are presented in Section~IV.
Finally, we provide a summary in Section~V.

\section{\label{sec:Model}Model and Formalism}

We consider a Hamiltonian that incorporates both tight-binding and interaction terms, expressed as follows:
\begin{equation}
	H =  \sum_{l,l'} \sum_{i,j,\tau,\tau',\sigma} t_{ij\tau\tau'}^{l,l'} c_{i\tau\sigma}^{l\dagger} c_{j\tau'\sigma}^{l'} 
	+ \sum_{l,i,\tau,\sigma} \varepsilon_l n_{i\tau\sigma}^l + H_I,
\end{equation}
where \( l/l' \), \( \tau/\tau' \), and \( \sigma \) denote the layer, orbital, and spin indices, respectively. The particle number operator is defined as \( n_{i\tau\sigma}^l = c_{i\tau\sigma}^{l\dagger} c_{i\tau\sigma}^l \). The layer-dependent onsite energy \( \varepsilon_l \) is introduced to account for the effects of an external perpendicular electric field.

The tight-binding parameters are adapted from Ref.~\cite{arXiv2501.06875} and adjusted to match experimental observations. Specifically, in Ref.~\cite{arXiv2501.06875}, a smaller interlayer hopping constant for the \(d_{z^2}\) orbital was proposed, resulting in weaker bonding-anti-bonding splitting. This leads to larger \(\gamma\) and \(\alpha\) Fermi surface pockets and the formation of an additional \(\delta\) pocket near the \(\Gamma\) point. 

However, recent ARPES experiments~\cite{arXiv2501.09255} measuring the normal-state Fermi surface revealed that the \(\delta\) pocket does not form, and the \(\gamma\) and \(\alpha\) pockets are smaller than predicted theoretically. These discrepancies suggest that the bonding-anti-bonding splitting may be stronger than previously estimated. 

In our present work, we adopt a larger interlayer hopping constant ($t^z_{\perp}$) to better align our model with ARPES results. Additionally, we adjust the chemical potential to ensure that the electron filling matches the experimentally measured value of 1.28~\cite{arXiv2501.09255}. We have numerically verified that our main conclusions are not sensitive to the value of $t^z_{\perp}$.
The tight-binding parameters of our model are summarized in Table~\ref{TABLES1}.

\begin{table}[h]
	\centering
	\begin{tabular}{|c|c|c|c|c|c|}
		\hline
		\multicolumn{1}{|c|}{$t_1^x$} & \multicolumn{1}{c|}{$t_1^z$} & \multicolumn{1}{c|}{$t_2^x$} & \multicolumn{1}{c|}{$t_2^z$} & \multicolumn{1}{c|}{$t_3^{xz}$}  & \multicolumn{1}{c|}{$t_{\perp}^x$}  \\ \hline
		-0.466 & -0.126 & 0.062 & -0.016 & 0.299 &0.001 \\ \hline
		$t_{\perp}^z$ & $t_{4}^{xz}$ & $\epsilon^x$ & $\epsilon^z$ & $t_{4}^x$ & $t_{4}^z$ \\ \hline
		-0.670  & -0.032 & 1.140  & 0.455 & -0.064 &-0.014  \\ \hline
		$t_{5}^x$ & $t_{5}^z$ & $t_{5}^{xz}$ &$t_{3}^x$ & $t_{3}^z$ & \\ \hline
		-0.015 & -0.003 & 0.026 & -0.001 & 0.033 & \\ \hline
	\end{tabular}
	\caption{\label{TABLES1} Tight-binding parameters of the bilayer two-orbital
		model.}
\end{table}

The interaction term $H_I$  is taken as the intra-orbital exchange interaction, given by:
\begin{equation}
	H_I =-\sum_{l,l'} \sum_{i,j,\tau}\sum_{\sigma,\sigma^\prime} V_{ij\tau}^{l,l'}  c_{i\tau\sigma}^{l\dagger} c_{j\tau\sigma^\prime}^{l'\dagger}c_{i\tau\sigma^\prime}^{l'} c_{j\tau\sigma}^{l} .
\end{equation}
At the mean-field level, the intralayer pairing order parameter is defined as
$\Delta_{ij\parallel}^{l\tau} = \frac{V_{\parallel}^{\tau}}{2} \left\langle c_{i\tau\uparrow}^l c_{j\tau\downarrow}^l - c_{i\tau\downarrow}^l c_{j\tau\uparrow}^l \right\rangle$,
where \( i \) and \( j \) denote nearest-neighbor sites within the same layer. For interlayer pairing, the interlayer interaction strength is given by \( V_{\perp}^{\tau} = V_{ii\tau}^{l,l'} \), and the corresponding mean-field order parameter is
$\Delta_{i\perp}^{\tau} = \frac{V_{\perp}^{\tau}}{2} \left\langle c_{i\tau\uparrow}^l c_{i\tau\downarrow}^{l'} - c_{i\tau\downarrow}^l c_{i\tau\uparrow}^{l'} \right\rangle$.

The Hamiltonian can be transformed into momentum space via a Fourier transformation, resulting in the expression
$H = \sum_{\bf k} \Psi^{\dagger}\left({\bf k}\right) \hat{M}\left({\bf k}\right) \Psi\left({\bf k}\right)$.
Here, \( \hat{M} \) is an \( 8 \times 8 \) matrix~\cite{supp}, and the Nambu spinor \( \Psi\left({\bf k}\right) \) is defined as
$    \Psi\left({\bf k}\right) = \left( u_{\bf k}, v_{\bf k} \right)^T$,
where
$u_{\bf k} = \left( c_{{\bf k}x\uparrow}^1, c_{{\bf k}z\uparrow}^1, c_{{\bf k}x\uparrow}^2, c_{{\bf k}z\uparrow}^2 \right)$,
and
$v_{\bf k} = \left( c_{-{\bf k}x\downarrow}^{1\dagger}, c_{-{\bf k}z\downarrow}^{1\dagger}, c_{-{\bf k}x\downarrow}^{2\dagger}, c_{-{\bf k}z\downarrow}^{2\dagger} \right)$.

The intralayer and interlayer pairing order parameters can be determined by
\begin{eqnarray}
	\Delta_{\parallel}^{l\tau} &=& \frac{1}{4N} \sum_{n{\bf k}} V_{\parallel}^{\tau} u_{nl\tau} v_{nl\tau}^* \tanh \left( \frac{\beta E_{n{\bf k}}}{2} \right) \left( \cos {\bf k_x} + \cos {\bf k_y} \right), \nonumber \\
	\\
	\Delta_{\perp}^{\tau} &=& \frac{1}{2N} \sum_{n{\bf k}} V_{\perp}^{\tau} u_{nl\tau} v_{nl'\tau}^* \tanh \left( \frac{\beta E_{n{\bf k}}}{2} \right).
\end{eqnarray}

The Green's function matrix is defined as:
\begin{equation}
	G_{ij}\left({\bf k},\omega\right) = \sum_n \frac{u_{in}\left({\bf k}\right) u^*_{jn}\left({\bf k}\right)}{\omega - E_n\left({\bf k}\right) + i\Gamma},
\end{equation}
where \( \Gamma \) is a damping factor, set to \( \Gamma = 0.002 \).

The spectral function of layer $l$ and orbital $\tau$ is expressed as:
\begin{equation}
	A^l_\tau\left({\bf k},\omega\right) = -\frac{1}{\pi} \mathrm{Im}  \left[ G_{pp}\left({\bf k},\omega\right) + G_{p+4,p+4}\left({\bf k},-\omega\right) \right],
\end{equation}
with \( p = \tau + 2(l - 1) \), where \( l \) and \( \tau \) take values \( 1 \) or \( 2 \).
The total spectral function is obtained by summing over the orbitals and the layers with $ A({\bf k},\omega)=\sum_{l,\tau} A^l_\tau({\bf k},\omega)$.

The density of states (DOS) as a function of frequency is calculated by integrating the spectral function over the entire Brillouin zone:
\begin{equation}
	\rho^l_\tau\left(\omega\right) = \frac{1}{N} \sum_{{\bf k}} A^l_\tau({\bf k},\omega).
\end{equation}
The DOS contribution from a specific orbital $\tau$ is then obtained by summing over the layer index, $\rho_\tau(\omega)=\sum_l \rho^l_\tau(\omega)$, while the total DOS is given by $\rho(\omega)=\sum_\tau \rho_\tau(\omega)$.

\section{Numerical results without Perpendicular Electric Fields }

Our investigation begins with an exploration of the pairing mechanism in Ruddlesden-Popper nickelate superconductors. The superconducting pairing strength $V$ is governed by the superexchange interaction $J$, where $J \approx 4t^2/U$. Consequently, it is widely accepted that the pairing strength $V$ is proportional to the square of the relevant hopping amplitude, i.e., $V \propto t^2$. In our model, the most dominant hopping parameter is the interlayer hopping for the $d_{z^2}$ orbital ($|t^z_{\perp}| = 0.670$), followed by the intralayer nearest-neighbor hopping for the $d_{x^2 - y^2}$ orbital ($|t^x_1| = 0.466$). Several studies have highlighted the significant role of Hund's coupling between the $d_{x^2 - y^2}$ and $d_{z^2}$ orbitals in enhancing superconducting pairing~\cite{PhysRevB.110.094509,PhysRevB.108.174511,PhysRevB.109.165154,PhysRevB.109.L081105,arXiv2311.12769}. Specifically, due to the substantial Hund's coupling, the $d_{z^2}$ and $d_{x^2 - y^2}$ orbitals can effectively share pairing strength with each other~\cite{PhysRevB.110.094509,PhysRevB.108.174511,PhysRevB.109.165154,PhysRevB.109.L081105,arXiv2311.12769}. Therefore, in the following analysis, we treat the pairing interactions for the $d_{x^2 - y^2}$ and $d_{z^2}$ orbitals on equal footing, setting $V_{\parallel}^x = V_{\parallel}^z = V_{\parallel}$ and $V_{\perp}^x = V_{\perp}^z = V_{\perp}$.

\begin{figure}
	\centering
	\includegraphics[width = 6cm]{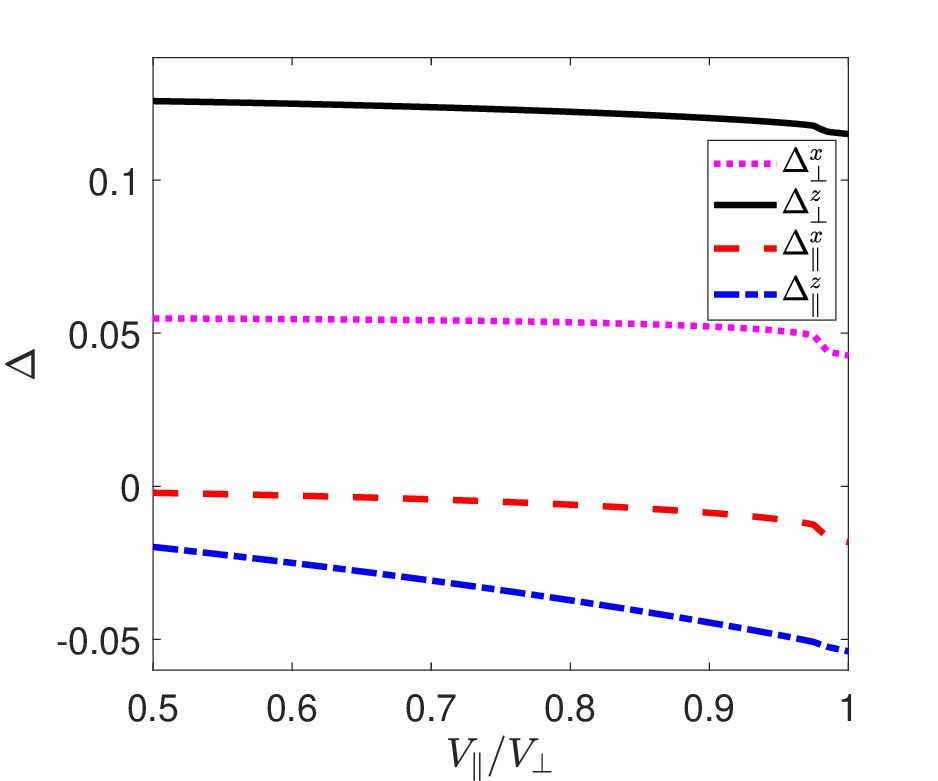}
	\caption{\label{fig:selfVd} Variation of the order parameters as a function of intralayer pairing strength \( V_{\parallel} \), ranging from 0.4 to 0.8, with the interlayer pairing strength fixed at  \( V_{\perp} = 0.8 \).}
\end{figure}

To investigate the role of intralayer pairing, we have performed a detailed study in which the interlayer pairing strength is fixed at $V_{\perp} = 0.8$, while the intralayer pairing strength $V_{\parallel}$ is varied from $0.4$ to $0.8$. The self-consistent results are presented in Fig.~\ref{fig:selfVd}. Within the considered parameter range, our results demonstrate that interlayer pairing is dominant. As the intralayer pairing potential increases, the amplitude of the intralayer $s$-wave pairing order parameter also increases, while the interlayer order parameter remains at a relatively high value. Even when the intralayer pairing potential equals the interlayer pairing potential, the largest pairing order parameter is still the interlayer pairing of the $d_{z^2}$ orbital, whereas the intralayer pairing of the $d_{x^2-y^2}$ orbital remains negligible. Additionally, it is noteworthy that the pairing signs of the intralayer and interlayer order parameters are always opposite.

Our self-consistent calculations indicate that the pairing magnitude in the $d_{x^2-y^2}$ orbital is much smaller than that in the $d_{z^2}$ orbital. Generally, the mean-field gap magnitudes are determined by the zero-energy DOS in the normal state. This result can thus be understood by examining the orbital-dependent normal-state DOS at the Fermi level~\cite{supp}. The zero-energy DOS of the $d_{z^2}$ orbital is much larger than that of the $d_{x^2-y^2}$ orbital, owing to the Van Hove singularity of the $\gamma$ band, which is primarily contributed by the $d_{z^2}$ orbital.

\begin{figure}
	\centering
	\includegraphics[width = 7cm]{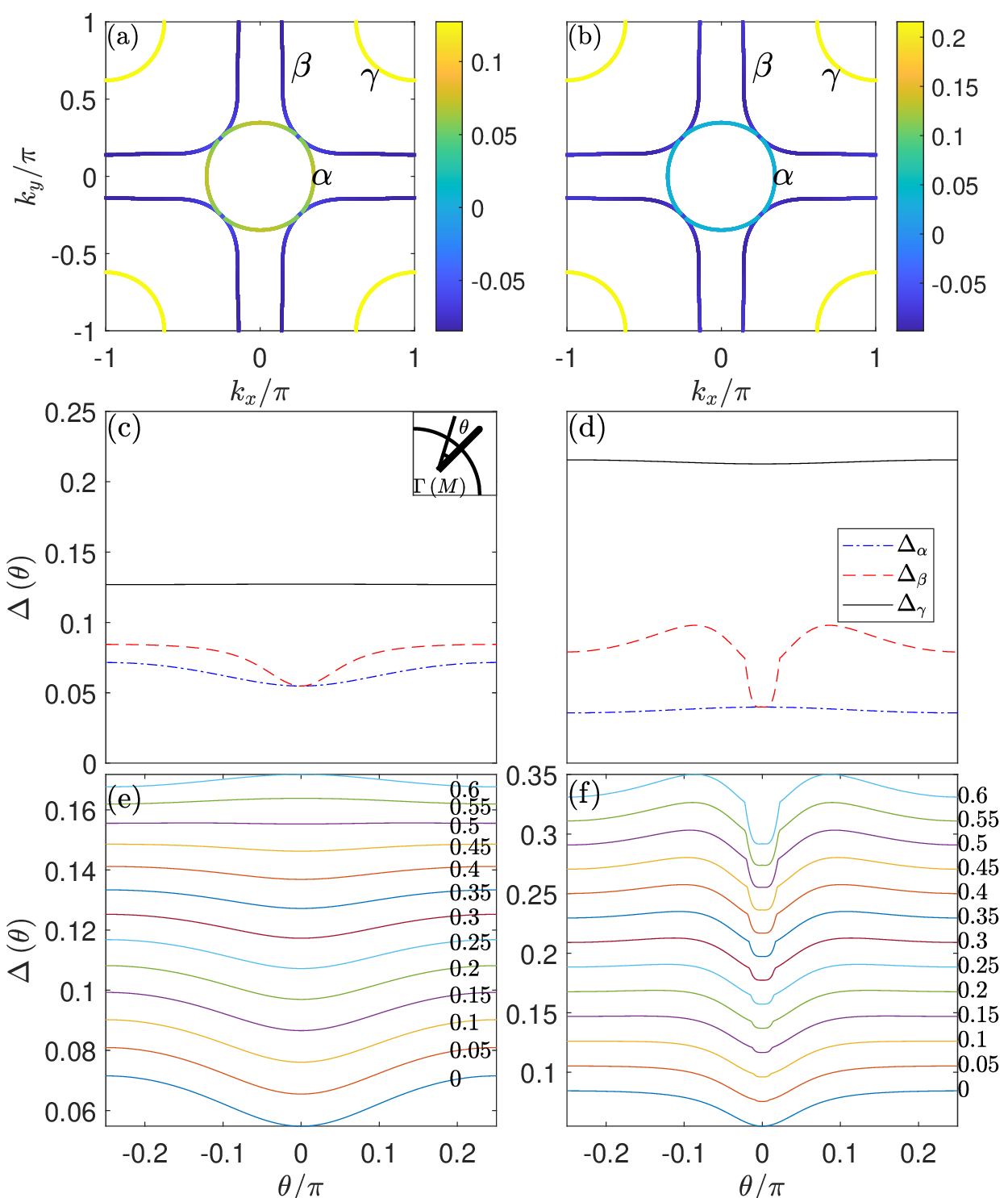}
	\caption{\label{fig:IBSgap} 
		(a) Gap function projected onto the Fermi surface with only interlayer pairing considered. (b) Similar to panel (a), but with coexistence of interlayer and intralayer pairing.
		(c) Magnitudes of the energy gap along different Fermi pockets with only interlayer pairing considered. (d) Similar to panel (c), but with coexistence of interlayer and intralayer pairing.	
		(e) Magnitudes of the energy gap along the \(\alpha\) Fermi pocket with the intralayer pairing strength \( V_{\parallel} \) increasing from 0 to 0.6. (d) Similar to panel (e), but for the \(\beta\) Fermi pocket.
	}
\end{figure}

We now explore the effective pairing structure along the Fermi surface pockets. The pairing function in the orbital basis can be transformed to the band basis via
\begin{equation}
	\tilde{\Delta}_{\mathbf{k}} = \Psi_0^{\dagger}(\mathbf{k})\, \Delta_{\mathbf{k}}\, \Psi_0(\mathbf{k}),
\end{equation}
where $\Psi_0(\mathbf{k})$ denotes the eigenvector matrix of the normal-state Hamiltonian, and $\Delta_{\mathbf{k}}$ represents the matrix form of the pairing term in the Hamiltonian~\cite{supp}. The four diagonal components of the pairing matrix $\tilde{\Delta}_{\mathbf{k}}$ correspond to the intraband pairings of the four normal-state bands, respectively.
These intraband pairings are the primary contributors to the superconducting gap. In contrast, the off-diagonal elements, representing interband pairings, approach zero at low energies and mainly influence the higher-energy excitation spectrum, without affecting the superconducting gap itself. Therefore, in the results presented below [Fig.~\ref{fig:IBSgap}], we focus exclusively on the intraband pairing components and neglect the effects of interband pairing.

Based on the mean-field self-consistent results shown in Fig.~\ref{fig:selfVd}, the intralayer $d_{x^2-y^2}$-wave pairing order parameter is negligible, prompting us to focus on purely interlayer pairing—a scenario supported by several theoretical studies~\cite{xie20243221,s41467-024-53863-5,PhysRevB.108.L201108,PhysRevB.108.174511,arXiv2311.05491,PhysRevLett.132.146002,arXiv2402.07449,yang2024decom,PhysRevB.111.094504}. 
We also examine a hybrid state combining interlayer and intralayer pairing, where interlayer pairing remains dominant, as indicated by mean-field calculations. Numerical results for the pairing order parameters along the normal-state Fermi surface are shown in Figs.~\ref{fig:IBSgap}(a) and \ref{fig:IBSgap}(b) for interaction configurations $(V_{\perp}, V_{\parallel}) = (0.8, 0)$ (pure interlayer coupling) and $(V_{\perp}, V_{\parallel}) = (0.8, 0.6)$ (mixed interlayer-intralayer coupling), respectively.

As shown, for both configurations, an effective $s_{\pm}$ pairing symmetry emerges: the pairing order parameter is positive along the $\gamma$ and $\alpha$ Fermi pockets and negative along the $\beta$ pocket. This characteristic $s_{\pm}$ pairing symmetry is consistent with previous theoretical results based on the spin fluctuation scenario~\cite{PhysRevLett.131.236002,PhysRevB.108.165141,PhysRevB.108.L140505,PhysRevLett.134.076001,PhysRevB.111.104505}.

We present the numerical results for the energy gap magnitudes along the three Fermi pockets for the two configurations in Figs.~\ref{fig:IBSgap}(c) and \ref{fig:IBSgap}(d), respectively. When only interlayer electron pairing is considered, as shown in Fig.~\ref{fig:IBSgap}(c), the superconducting gap is nodeless along all three Fermi pockets. Along the $\gamma$ Fermi pocket, the gap remains constant ($\Delta_\gamma = 0.13$). For the $\alpha$ and $\beta$ Fermi pockets, the gap is anisotropic, reaching a minimum along the diagonal direction ($\theta = 0$) and increasing monotonically with $\theta$. This behavior is consistent with recent ARPES experimental results on the gap magnitude near the diagonal direction~\cite{arXiv2502.17831}.

When both interlayer and intralayer electron pairing are present, as depicted in Fig.~\ref{fig:IBSgap}(d), the gap magnitude along the $\gamma$ Fermi pocket increases significantly ($\Delta_\gamma \approx 0.21$). The gap behavior along the $\alpha$ and $\beta$ Fermi surfaces differs markedly from the case with only interlayer pairing. Along the $\alpha$ pocket, the gap decreases slightly with increasing $\theta$, while along the $\beta$ pocket, the gap exhibits non-monotonic behavior, first increasing and then decreasing as a function of $\theta$.

The enhancement of the gap magnitude along the $\gamma$ pocket in the presence of intralayer pairing can be understood by analyzing the intralayer pairing function. The intralayer order parameter has a negative sign [Fig.~\ref{fig:selfVd}], while the $\gamma$ Fermi pocket is located near the $M$ point ($M = (\pi, \pi)$), where the intralayer pairing generates an effective positive gap that constructively combines with the interlayer component. For the $\alpha$ and $\beta$ pockets, near the diagonal direction, the intralayer pairing introduces a negative gap contribution, while the interlayer pairing generates a positive gap along the $\alpha$ pocket and a negative gap along the $\beta$ pocket [Fig.~\ref{fig:IBSgap}(a)]. As a result, for small $\theta$, the pairing gap along the $\beta$ pocket is enhanced, whereas the gap along the $\alpha$ pocket is suppressed. For larger values of $\theta$, the gap magnitude along the $\beta$ pocket decreases because the intralayer pairing generates a positive gap contribution when crossing the nodal point of the intralayer $s$-wave pairing function.

We further examine the impact of the intralayer pairing component on the superconducting gap magnitude of the $\alpha$ and $\beta$ Fermi pockets. Figs.~\ref{fig:IBSgap}(e) and \ref{fig:IBSgap}(f) show the $\theta$-dependence of the gap along the $\alpha$ and $\beta$ pockets, respectively, with $V_\perp = 0.8$ and $V_\parallel$ increasing from 0 to 0.6. Along the $\alpha$ pocket [Fig.~\ref{fig:IBSgap}(e)], the gap increases monotonically with $\theta$ when $V_{\parallel} < 0.5$, but decreases monotonically when $V_{\parallel} > 0.5$. For the $\beta$ pocket [Fig.~\ref{fig:IBSgap}(f)], the gap magnitude exhibits pronounced variations near the diagonal direction as $V_{\parallel}$ increases. Notably, when $V_{\parallel} > 0.25$, the gap develops non-monotonic behavior, first increasing and then decreasing with $\theta$. These gap variations can be detected in future experiments to determine the relative weight of the intralayer pairing component.

Our results show that all three Fermi pockets are fully gapped when interlayer $s$-wave pairing is dominant, in agreement with recent experimental findings~\cite{arXiv2506.01788,arXiv2502.17831}. The resulting spectral function and gap structure differ significantly from those associated with pure 
$d_{x^2-y^2}$-wave pairing symmetry, which features nodal points along the diagonal direction. These distinctions can be clearly resolved by ARPES and STM experiments. Moreover, our self-consistent calculations do not support the coexistence of 
$s$-wave and 
$d$-wave pairing states, which is consistent with the symmetry analysis presented in Ref.~\cite{arXiv2506.01788}.

For pure interlayer pairing, the energy gap magnitudes along the $\alpha$ and $\beta$ Fermi pockets are comparable and exhibit a slight increase with $\theta$, consistent with recent experimental observations~\cite{arXiv2502.17831}. However, current experimental data provide gap information for the $\alpha$ and $\beta$ pockets only within a very narrow range near the diagonal, which limits further theoretical and experimental comparisons.

\section{Numerical Results with Additional Perpendicular Electric Fields}

As mentioned in Sec. I, bilayer nickelate superconductors feature two electronically equivalent NiO$_2$ layers. If this equivalence between the two layers is experimentally disrupted, interlayer pairing may induce nodes or nodal lines, providing a critical test for the dominance of interlayer pairing in these materials. In practice, one can introduce an additional effective perpendicular electric field to break this equivalence.

Theoretically, we model this effect by introducing a layer-dependent chemical potential term in Eq.~(1), expressed as $\varepsilon_l = \frac{(-1)^l V_F}{2}$, where $V_F$ denotes the interlayer voltage. This term explicitly breaks the equivalence between the two layers and simulates the effect of a perpendicular electric field.

In the following, we consider two distinct cases in our numerical results. In Sec.~IV(A), we focus on the case of pure interlayer pairing by setting the intralayer pairing strength to zero, i.e., $(V_{\perp}, V_{\parallel}) = (0.8, 0)$. In Sec.~IV(B), we examine the coexistence of interlayer and intralayer pairing, with $(V_{\perp}, V_{\parallel}) = (0.8, 0.6)$.

\begin{figure}
	\centering
	\includegraphics[width=8cm]{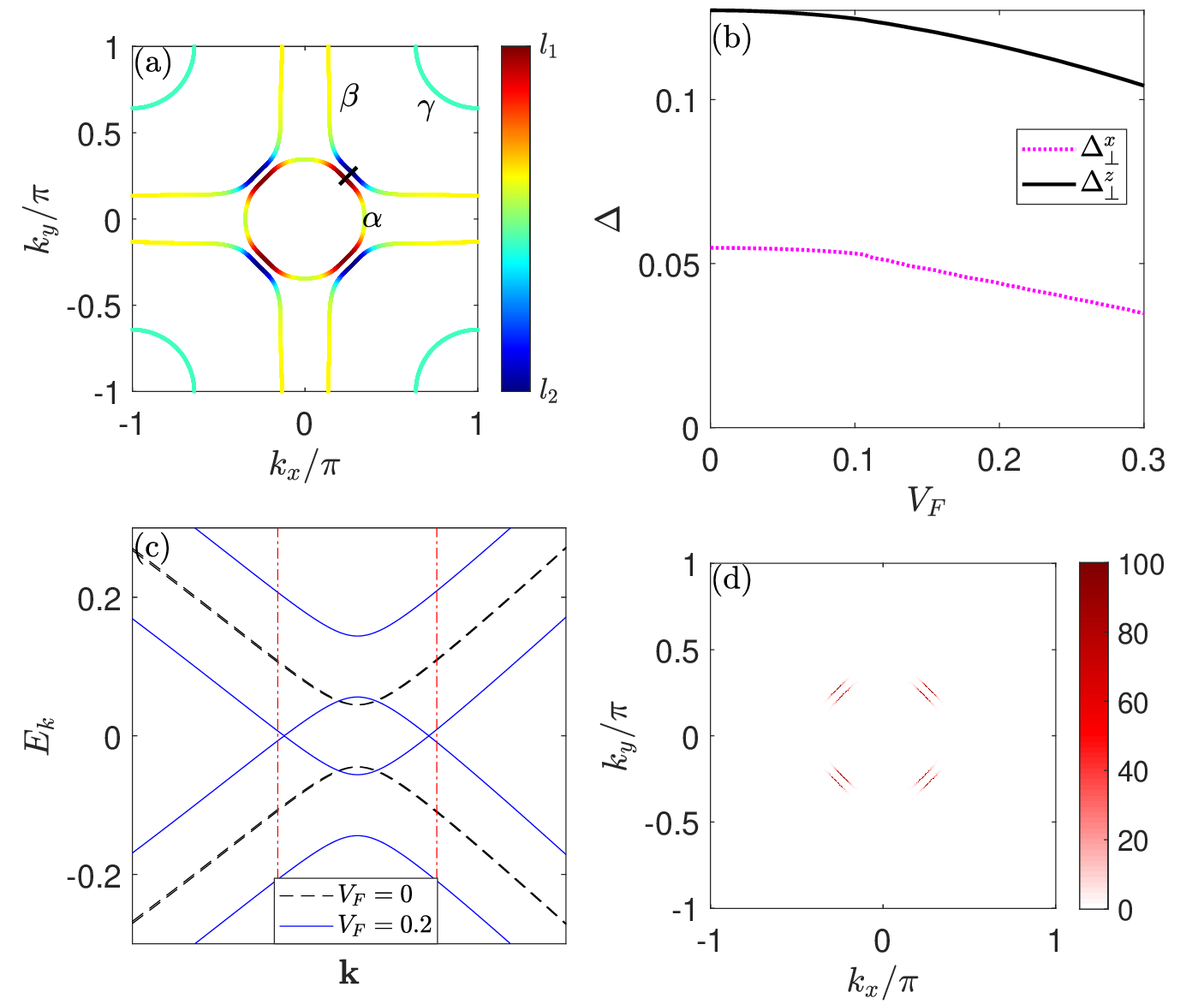}
	\caption{\label{fig3} (a) Normal-state Fermi surface with an applied interlayer voltage $ V_F = 0.2 $. (b) Self-consistent results of interlayer pairing order parameters as a function of voltage $ V_F $.
		(c) Superconducting energy bands along the short line cut in panel (a), comparing $ V_F = 0.0 $ (black dashed lines) and $ V_F = 0.2 $ (blue solid lines). The (red) dash-dotted lines indicate the position of the normal state Fermi surface.   
		(d) Zero-energy total spectral function in the superconducting state with $ V_F = 0.2 $.}
\end{figure}

\subsection{Results with Pure Interlayer Pairing}

We replot the normal-state Fermi surface in the presence of an additional effective perpendicular electric field with $V_F = 0.2$ in Fig.~\ref{fig3}(a). As shown, the application of a perpendicular electric field induces significant modifications to the $\alpha$ and $\beta$ Fermi pockets near the diagonal direction. In the absence of the field, these two pockets are adjacent along the diagonal. Under the electric field, however, the $\alpha$ and $\beta$ pockets become separated. Furthermore, near the diagonal, the $\alpha$ pocket is dominated by quasiparticles from layer 1, while the $\beta$ pocket primarily originates from layer 2.

This pronounced layer inequivalence near the diagonal direction arises from the vanishing inter-orbital coupling and the weak interlayer coupling of the $d_{x^2-y^2}$ orbital in this region. As a result, the $d_{x^2-y^2}$ orbitals in the two layers become effectively decoupled along the diagonal, allowing the electric field to easily split the doubly degenerate bands into two bands dominated by different layers. Since the formation of an interlayer Cooper pair requires one electron from each layer, this splitting has important consequences for superconductivity.

In Fig.~\ref{fig3}(b), we present self-consistent calculations of the interlayer pairing order parameters as a function of the voltage $V_F$. For both orbitals, the interlayer pairing order parameter decreases with increasing $V_F$, consistent with the broken equivalence between the two NiO$_2$ layers. Notably, even under a strong field ($V_F = 0.3$), the interlayer pairing in the $d_{z^2}$ orbital retains a relatively large value (about 0.1).

Fig.~\ref{fig3}(c) displays the energy bands in the superconducting state near the $\alpha$ and $\beta$ Fermi pockets along the diagonal direction (along the short line in Fig.~3(a)). At $V_F=0$, the $\alpha$ and $\beta$ bands are degenerate and exhibit a full superconducting gap. When $V_F=0.2$, however, the bands split, generating two nodes at the $\alpha$ and $\beta$ Fermi pockets, respectively. It is worth noting that the positions of these nodes are slightly shifted from those of the normal-state Fermi surface.

Fig.~\ref{fig3}(d) illustrates the zero-energy total spectral function of the superconducting state, $A({\bf k},\omega=0)$, at $V_F=0.2$, revealing two disconnected Fermi arcs. These arcs arise from the two electronically inequivalent NiO$_2$ layers, which cannot form Cooper pairs due to their electronic inequivalence [see Fig.~3(a)]. The arc size depends sensitively on $V_F$. Our calculations show that the arcs first emerge at $V_F=0.11$ and grow progressively with increasing $V_F$~\cite{supp}.

\begin{figure}
	\centering
	\includegraphics[width = 8cm]{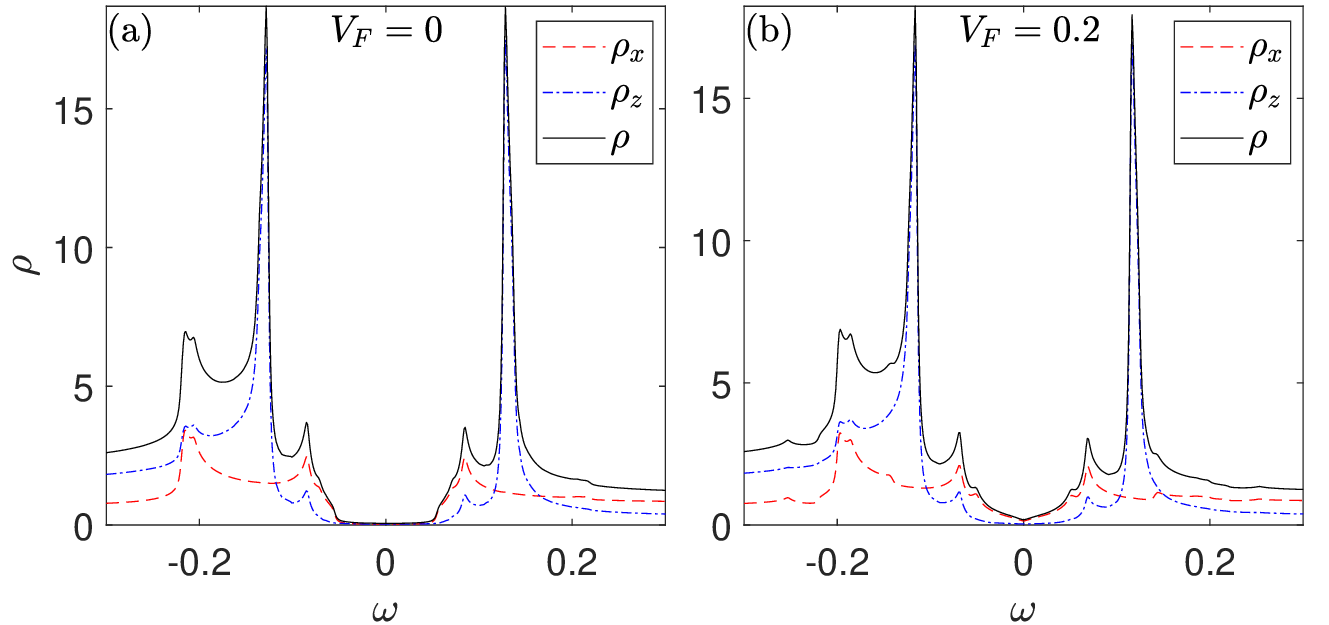}
	\caption{\label{fig4} DOS spectra in the superconducting state considering the pure interlayer pairing with (a) \( V_F = 0 \) and (b) \( V_F = 0.2 \).  }
\end{figure}

We now discuss in more detail the origin of the emergence of Fermi arcs. Along the diagonal direction, the normal-state $\alpha$ and $\beta$ Fermi pockets are entirely contributed by the $d_{x^2-y^2}$ orbital~\cite{supp}. In this case, the superconducting Hamiltonian (presented in the supplementary material~\cite{supp}) can be simplified to a $4 \times 4$ matrix by neglecting interorbital coupling. Furthermore, for the $d_{x^2-y^2}$ orbital, the interlayer hopping constant is rather small, so the normal-state band structures of the two layers can be regarded as independent. As a result, in the normal state, in the presence of an additional perpendicular electric field, the normal-state Fermi surface pockets near the diagonal direction become separated and originate from different layers. Consequently, intralayer pairing leads to intraband pairing, while interlayer pairing leads to interband pairing. Here, we focus on the case of pure interlayer pairing. As previously mentioned, the effective interband pairing at the Fermi level is nearly zero, and as a result, Fermi arcs emerge.

The above qualitative discussion can be substantiated by further analytical calculations. Without the external field, the superconducting quasiparticle energy bands along the diagonal direction are degenerate due to the equivalence of the two layers. With pure interlayer pairing, the quasiparticle energy for the $d_{x^2-y^2}$ orbital is given by
$E_{\mathbf{k}} = \pm \sqrt{(T_{\mathbf{k}}^x)^2 + (\Delta_{\perp}^x)^2}$,
where $T_{\mathbf{k}}^x$ represents the normal-state quasiparticle energy~\cite{supp}. In this case, the energy bands are fully gapped, with the minimum excitation energy being $\Delta_{\perp}^x$, as shown in Fig.~\ref{fig:IBSgap}.

In the presence of an additional perpendicular electric field, the $d_{x^2-y^2}$ orbital quasiparticle energy along the diagonal direction is obtained by diagonalizing the normal-state Hamiltonian, resulting in
\begin{equation}
	E_{n\mathbf{k}} = \pm \left( \frac{V_F}{2} \pm \sqrt{(T_{\mathbf{k}}^x)^2 + (\Delta_{\perp}^x)^2} \right).
\end{equation}
Nodes emerge when $V_F \geq 2\Delta_{\perp}^x$, which naturally explains the appearance of Fermi arcs when $V_F \geq 0.11$.

It is important to note that, in the superconducting state, the node positions along the diagonal direction are determined by the condition
\begin{equation}
	\frac{V_F}{2} = \sqrt{(T_{\mathbf{k}}^x)^2 + (\Delta_{\perp}^x)^2},
\end{equation}
with the node positions deviating slightly from the normal-state Fermi surface.

We now analyze the modifications to the DOS spectrum induced by an external electric field. Fig.~\ref{fig4} compares the calculated DOS for both field-free and field-applied configurations. In the absence of an external electric field ($V_F = 0$), the system is fully gapped, and both the $d_{x^2-y^2}$ and $d_{z^2}$ orbitals exhibit U-shaped spectral profiles. The coherent peak associated with the $d_{z^2}$ orbital appears at higher energies and with enhanced intensity.

Upon applying an external electric field ($V_F = 0.2$), the energy gap of the $d_{z^2}$ orbital is reduced, but its DOS near the Fermi level remains negligible. In contrast, the $d_{x^2-y^2}$ orbital displays the emergence of finite low-energy states, indicative of unpaired quasiparticles. These results are consistent with the nodal structure observed in the superconducting state (see Fig.~\ref{fig3}), underscoring the critical role of layer inequivalence in modulating the energy gap under external fields.

\begin{figure}
	\centering
	\includegraphics[width = 8cm]{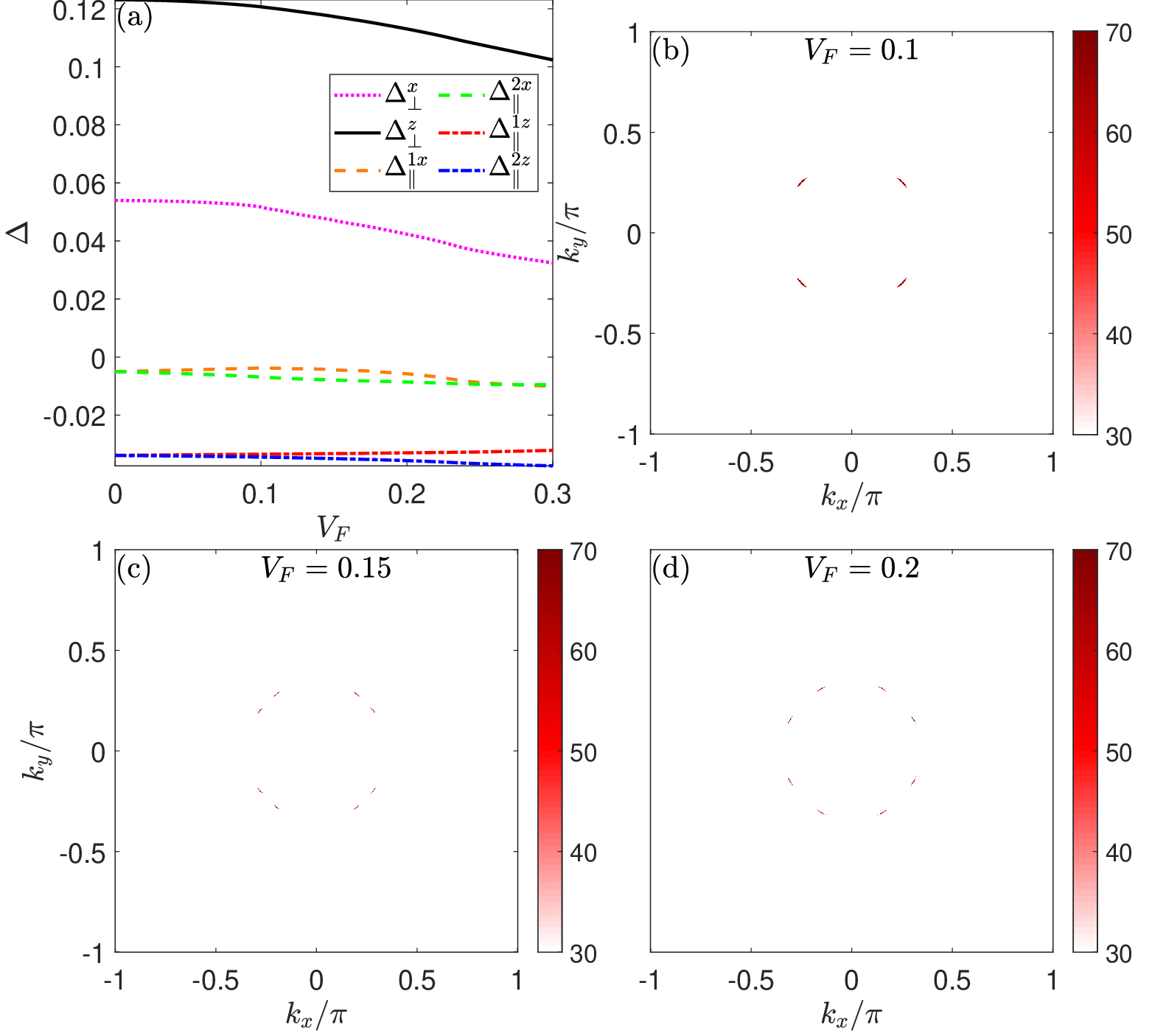}
	\caption{\label{fig5} (a) Self-consistent interlayer and intralayer pairing order parameters as a function of the applied voltage $V_F$. (b-d) Zero-energy spectral function in the superconducting state, considering the coexistence of interlayer and intralayer pairing for different interlayer voltages $V_F$. }
\end{figure}

\subsection{Results Considering the Coexistence of Interlayer and Intralayer Pairing}

We now consider the coexistence of intralayer and interlayer pairing. The self-consistent results for the intralayer and interlayer pairing gaps as functions of the interlayer voltage $V_F$ are shown in Fig.~\ref{fig5}(a). As can be seen, the applied electric field breaks the equivalence between layers, resulting in different intralayer order parameters for the two layers. The intralayer order parameters exhibit only a weak dependence on $V_F$, while the interlayer order parameters decrease slightly as $V_F$ increases. Notably, even for a relatively large voltage ($V_F = 0.3$), the interlayer pairing order parameter for the $d_{z^2}$ orbital remains dominant and significantly larger than the pairing parameters in other channels.

We present the zero-energy spectral function for different values of $V_F$ in Figs.~\ref{fig5}(b)-\ref{fig5}(d). As shown, these results are markedly different from the pure interlayer case. Here, nodal points appear in the $\alpha$ pocket for various values of $V_F$, while the $\beta$ pocket remains fully gapped. The emergence of these nodal points in the $\alpha$ pocket under an applied electric field may serve as a useful experimental signature for probing the coexistence of interlayer and intralayer pairing.

These results can be understood by analyzing the mean-field pairing order parameters. As shown in Fig.~\ref{fig5}(a), the intralayer order parameters are negative. This generates a positive pairing gap on the $\gamma$ Fermi pocket, thereby enhancing the maximum pairing gap. Conversely, the negative intralayer pairing order parameters induce negative superconducting gaps near the diagonal directions of the $\alpha$ and $\beta$ pockets. The interlayer pairing produces a negative gap along the $\beta$ pocket and a positive gap along the $\alpha$ pocket. In the absence of an electric field, interlayer pairing dominates, resulting in a nodeless energy gap across the entire Fermi surface. However, in the presence of an electric field, as discussed above, the gap induced by interlayer pairing near the diagonal directions of the $\alpha$ and $\beta$ pockets is significantly suppressed. For the $\alpha$ pocket, the negative gap from intralayer pairing and the positive gap from interlayer pairing coexist and may compete, leading to nodal points where the gaps cancel each other. For the $\beta$ pocket, both interlayer and intralayer pairings contribute negative gaps, so no nodal points exist in this case.

\begin{figure}
	\centering
	\includegraphics[width = 8cm]{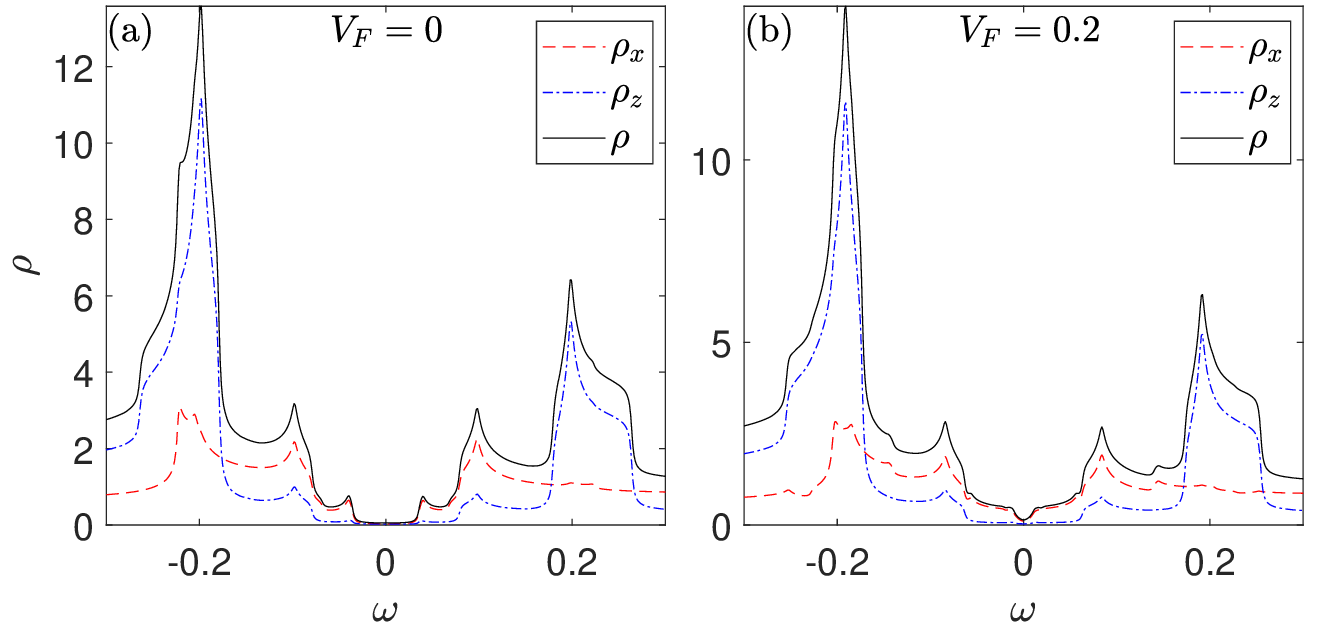}
	\caption{\label{fig6} DOS spectra in the superconducting state with coexisting interlayer and intralayer pairing for (a) $V_F = 0$ and (b) $V_F = 0.2$.  }
\end{figure}

The existence of nodal points can also be detected through the DOS spectra, as shown in Fig.~\ref{fig6}. In Fig.~\ref{fig6}(a), the absence of an electric field results in a fully gapped system with a U-shaped gap. In contrast, Fig.~\ref{fig6}(b) shows a V-shaped gap in the presence of an electric field, indicating the presence of nodal points.

Finally, we discuss possible approaches to realize an effective perpendicular electric field, which is essential for our proposal. In practice, a perpendicular electric field in thin-film systems can be effectively achieved and well controlled using a double-gate device~\cite{Oostinga2008,Zhang2009}. This technique has been widely employed to realize various exotic quantum states in bilayer graphene systems~\cite{doi:10.1126/science.1208683,science.abm8386,science.adk9749,s41586-024-07584-w,PhysRevLett.133.066301}. In the system considered here, the distance between the two NiO$_2$ layers is approximately 
4.28~\AA~\cite{arXiv2501.06875}. In our calculations, Fermi arcs or nodal points emerge when the interlayer voltage exceeds $0.11$, corresponding to a displacement field strength of about 
$0.26\varepsilon$~V/nm, where $\varepsilon$
is the 
$c$-axis dielectric constant. For layered materials, the 
$c$-axis dielectric constant is generally much smaller than the in-plane value, typically ranging from
$5$ to $10$~\cite{Laturia2018}. With the double-gate technique, the displacement field can reach as high as 
$3$~V/nm~\cite{Zhang2009}. Therefore, the parameters used in our proposal are reasonable and should be accessible experimentally.

An advantage of the double-gate technique is that both the Fermi level of each layer and the electric field strength can be precisely controlled. However, the presence of a gate device may affect ARPES measurements. As an alternative, an effective electric field can also be induced by chemical techniques, such as interface doping~\cite{nmat3439,NR.2025.94907360} or inserting additional atomic layers~\cite{PMID26351697,Chapman2016,Ichinokura2016}, which can also break layer equivalence. Such chemical approaches are compatible with ARPES measurements, but achieving precise control of the field strength is more challenging.

\section{summary}
In summary,
recent ARPES experiments have measured the superconducting gap magnitude of La$_{2.85}$Pr$_{0.15}$Ni$_2$O$_7$ thin films in limited regions of the Brillouin zone, marking the first direct observation of the superconducting gap in Ruddlesden-Popper nickelate superconductors. Our calculations show that the gap magnitudes resulting from interlayer pairing are in qualitative agreement with these experimental results within the measured region. However, due to experimental constraints, the available gap information is restricted to a very narrow portion of the Brillouin zone.

Our research suggests that the application of external electric fields could make highly efficient use of this limited experimental information to probe the underlying pairing mechanisms. Such electric fields can be effectively realized through gate tuning, interface doping, or by inserting additional atomic layers. In the scenario with only interlayer pairing, Fermi arcs emerge when the interlayer voltage $V_F$
exceeds a critical value $V^c_F=0.11$. When both interlayer and intralayer pairing coexist, nodal points can be induced for $V_F\geq 0.1$. These nodal points appear exclusively on the 
pocket, while the 
pocket remains fully gapped. Our numerical results thus provide a promising avenue to probe the pairing mechanisms in La$_{2.85}$Pr$_{0.15}$Ni$_2$O$_7$ thin films.




%

\renewcommand{\thesection}{S-\arabic{section}}
\setcounter{section}{0}  
\renewcommand{\theequation}{S\arabic{equation}}
\setcounter{equation}{0}  
\renewcommand{\thefigure}{S\arabic{figure}}
\setcounter{figure}{0}  
\renewcommand{\thetable}{S\Roman{table}}
\setcounter{table}{0}  
\onecolumngrid \flushbottom 
\newpage
\begin{center}\large \textbf{Supplemental material for effective perpendicular electric field as a probe for interlayer pairing in ambient-pressure superconducting La$_{2.85}$Pr$_{0.15}$Ni$_{2}$O$_{7}$ thin films} \end{center}

\section{tight-binding model}
In momentum space,
the Hamiltonian can be expressed as $H = \sum_{\bf k} \Psi_k^{\dagger} H\left({\bf k}\right) \Psi_k$, where the basis vector is defined as $\Psi_k^{\dagger} = \left( c_{{\bf k}x\uparrow}^{1\dagger}, c_{{\bf k}z\uparrow}^{1\dagger}, c_{{\bf k}x\uparrow}^{2\dagger}, c_{{\bf k}z\uparrow}^{2\dagger}, c_{-{\bf k}x\downarrow}^1, c_{-{\bf k}z\downarrow}^1, c_{-{\bf k}x\downarrow}^2, c_{-{\bf k}z\downarrow}^2 \right)$. 

The Hamiltonian matrix \(H({\bf k})\) is an \(8 \times 8\) matrix,
\begin{eqnarray}
	H\left({\bf k}\right) = \left( \begin{array}{cc}
		{H_t\left({\bf k}\right)}&{H_{\Delta}\left({\bf k}\right)}\\
		{H_{\Delta}^{\dagger}\left({\bf k}\right)}&{-H_t\left({\bf k}\right)}
	\end{array} \right),
\end{eqnarray}
where \(H_t({\bf k})\) represents the normal-state tight-binding Hamiltonian in a \(4 \times 4\) matrix form,
\begin{eqnarray}
	H_{t}\left({\bf k}\right) = \left( \begin{array}{cc}
		{H_A\left({\bf k}\right)}&{H_{AB}\left({\bf k}\right)}\\
		{H_{AB}\left({\bf k}\right)}&{H_A\left({\bf k}\right)}
	\end{array} \right).
\end{eqnarray}
The matrices \(H_A({\bf k})\) and \(H_{AB}({\bf k})\) are given by
\begin{eqnarray}
	H_A\left({\bf k}\right) = \left( \begin{array}{cc}
		{T_{\bf k}^x}&{V_{\bf k}}\\
		{V_{\bf k}}&{T_{\bf k}^{z}}
	\end{array} \right) 
	,
	H_{AB}\left({\bf k}\right) = \left( \begin{array}{cc}
		{{T'}_{\bf k}^x}&{V'_{\bf k}}\\
		{V'_{\bf k}}&{{T'}_{\bf k}^z}
	\end{array} \right).
\end{eqnarray}
The components are defined as
\begin{eqnarray}
	T_{\bf k}^{x/z} &=& 2t_{1}^{x/z}\left( \cos { k_x} + \cos { k_y} \right) + 4t_{2}^{x/z} \cos { k_x} \cos { k_y} + \nonumber \\
	&& 2t_{4}^{x/z}\left( \cos 2{ k_x} + \cos 2{ k_y} \right) + 2t_{5}^{x/z}\left( \cos 3{ k_x} + \cos 3{ k_y} \right) + \epsilon_{x/z}, \\
	{T'}_{\bf k}^{x/z} &=&  t_{\perp}^{x/z} + 2t_3^{x/z} \left( \cos { k_x} + \cos { k_y} \right), \\
	V_{\bf k} &=& 2t_{3xz}\left( \cos { k_x} - \cos { k_y} \right), \\
	V'_{\bf k} &=& 2t_{4xz}\left( \cos { k_x} - \cos { k_y} \right).
\end{eqnarray}

The superconducting pairing part of the Hamiltonian, \(H_\Delta({\bf k})\), takes the form
\begin{eqnarray}
	H_{\Delta}\left({\bf k}\right) = \left( \begin{array}{cccc}
		{\Delta_{{\bf k} \parallel}^{1x}}&{0}&{\Delta_{\perp}^x}&{0}\\
		{0}&{\Delta_{{\bf k} \parallel}^{1z}}&{0}&{\Delta_{\perp}^z}\\
		{\Delta_{\perp}^x}&{0}&{\Delta_{{\bf k} \parallel}^{2x}}&{0}\\
		{0}&{\Delta_{\perp}^z}&{0}&{\Delta_{{\bf k} \parallel}^{2z}}\\
	\end{array} \right),
\end{eqnarray}
where $\Delta_{{\bf k} \parallel}^{l\tau} = 2\Delta_{\parallel}^{l\tau} \left( \cos { k_x} + \cos { k_y} \right)$. $\Delta_{\parallel}^{l\tau}$ and $\Delta_{\perp}^\tau$ denote the intralayer and interlayer superconducting order parameters, respectively, which are determined self-consistently.

\newpage

\section{Normal-state Electronic structure}

\renewcommand \thefigure {S\arabic{figure}}
\begin{figure}[h]
	\centering
	\includegraphics[width = 12cm]{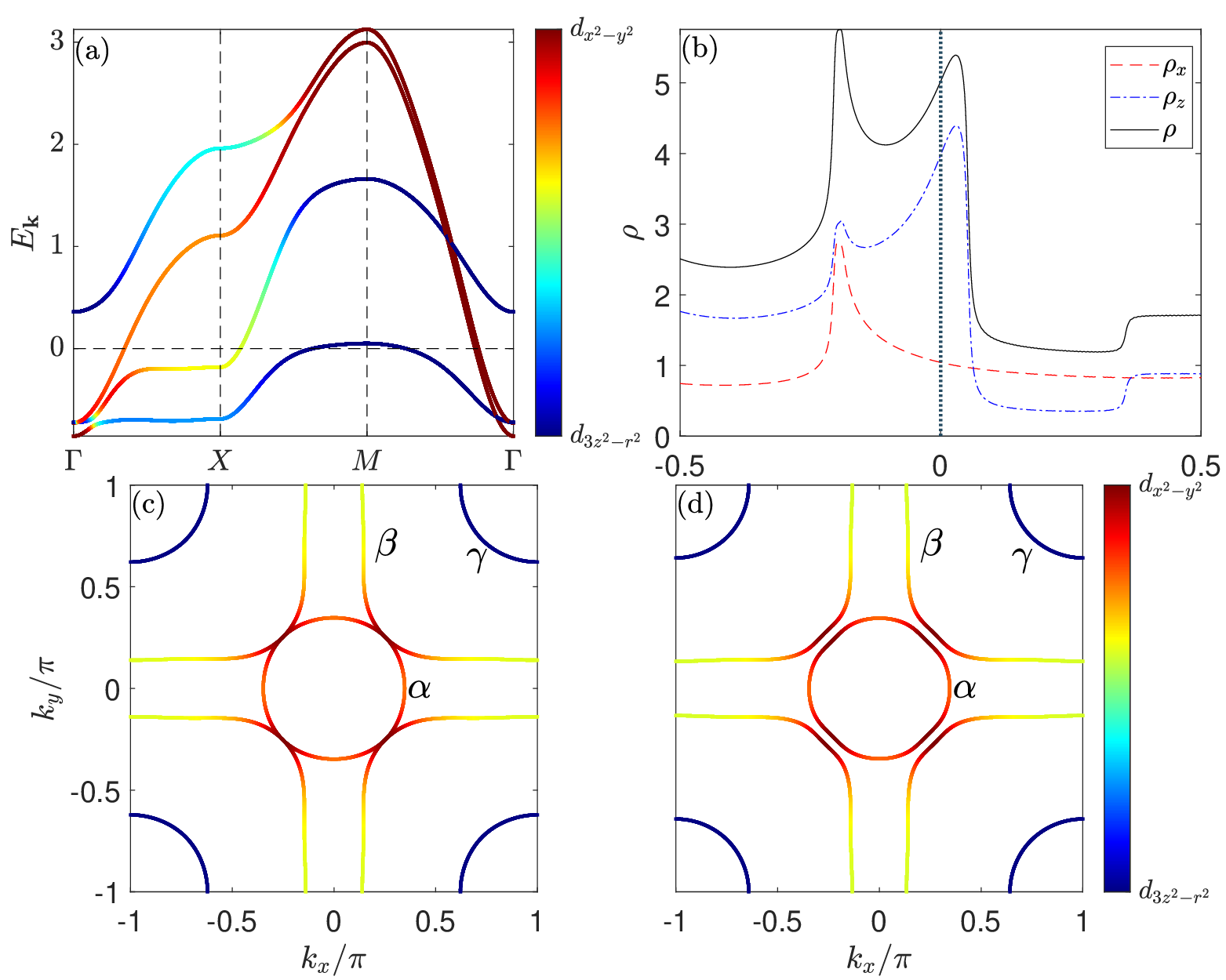}
	\caption{\label{figS1}Numerical results for the normal-state electronic structure. (a) Normal-state energy bands in the absence of an external field. (b) Orbital-resolved and total density of states without the external field. Panels (c) and (d) display normal-state Fermi surfaces for 
		$V_F=0$ and $V_F=0.2$, respectively. The colors in panels (a), (c), and (d) represent the orbital weights.
	}
\end{figure}

The superconducting properties presented in the main text can be preliminarily understood by exploring the normal-state electronic structure. Previously, the normal-state electronic structure of pressurized bulk La$_3$Ni$_2$O$_7$ has been studied using a two-orbital model~\cite{PhysRevLett.131.126001}. Here, we reexamine the normal-state structure based on a similar two-orbital model, but with the parameters given in the main text, which are more suitable for La$_{2.85}$Pr$_{0.15}$Ni$_2$O$_7$ thin films~\cite{arXiv2501.06875,arXiv2501.09255}. The normal-state energy bands, obtained by diagonalizing the Hamiltonian in Eq.~S2, are shown in Fig.~\ref{figS1}(a). The orbital-resolved and total density of states (DOS) spectra are presented in Fig.~\ref{figS1}(b).

As shown in Fig.~\ref{figS1}(a), the normal-state energy bands exhibit a $d_{z^2}$-dominant flat band near the $M$ point (Van Hove singularity) at a low positive energy, resulting in a larger DOS of the $d_{z^2}$ orbital at the Fermi level, as shown in Fig.~\ref{figS1}(b). At the mean-field level, a larger DOS at the Fermi level generally leads to a larger superconducting gap magnitude. As a result, the superconducting order parameters for the $d_{x^2-y^2}$ orbital are significantly smaller than those for the $d_{z^2}$ orbital, as presented in Fig.~1 of the main text.

The normal-state Fermi surfaces for $V_F=0$ and $V_F=0.2$ are shown in Figs.~\ref{figS1}(c) and \ref{figS1}(d), respectively. In both cases, the $\gamma$ pocket is primarily contributed by the $d_{z^2}$ orbital. The $\alpha$ and $\beta$ pockets receive contributions from both orbitals; however, near the diagonal direction, these two pockets are mainly dominated by the $d_{x^2-y^2}$ orbital. These features are consistent with the band structure displayed in Fig.~\ref{figS1}(a) and are qualitatively similar to those obtained from previous two-orbital models~\cite{PhysRevLett.131.126001}. Understanding these characteristics is important for elucidating the origin of the Fermi arcs induced by the electric field, as discussed in the main text.

\section{Zero Energy Spectral Function in the Superconducting State with Only the Interlayer Pairing in the Presence of an Additional Electric Field}
\renewcommand \thefigure {S\arabic{figure}}
\begin{figure}[h]
	\centering
	\includegraphics[width = 18cm]{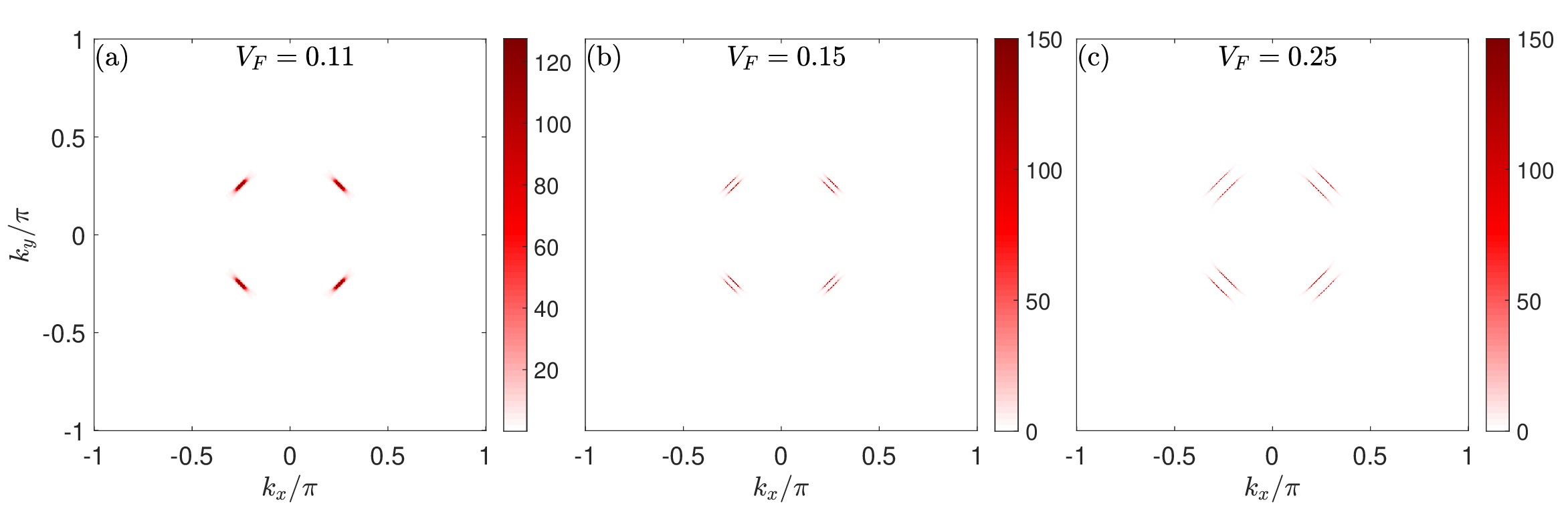}
	\caption{\label{figS3} Zero-energy spectral function in the superconducting state where only the interlayer pairing exists with (a) $V_F = 0.11$, (b) $V_F = 0.15$ and (c) $V_F = 0.25$.}
\end{figure}

In the main text, we investigate the effect of an additional electric field on the interlayer pairing. Specifically, we consider an interlayer voltage \(V_F = 0.2\) to demonstrate the emergence of Fermi arcs in the superconducting state [see Fig. 3(d) in the main text]. Here, we present additional numerical results for the zero-energy spectral function across different voltage values in Fig.~\ref{figS3}.

Our calculations reveal that Fermi arcs first emerge at \(V_F = 0.11\), coinciding with the initial separation of the \(\alpha\) and \(\beta\) pockets. As \(V_F\) increases further, as shown in Fig.~\ref{figS3}(b) and \ref{figS3}(c), the size of the Fermi arcs gradually expands. Additionally, the two arcs originating from the \(\alpha\) and \(\beta\) pockets become increasingly well-separated with increasing voltage.


\end{document}